\documentclass[aps,prb,twocolumn,groupedaddress,showpacs]{revtex4}
\usepackage{graphicx}
\usepackage{amsmath,amssymb,amsfonts}
\usepackage{natbib}

\def\beq{\begin{equation}}
\def\eeq{\end{equation}}
\def\bdm{\begin{displaymath}}
\def\edm{\end{displaymath}}
\newcommand {\ket}[1]{|{#1}\rangle}
\newcommand {\brak}[1]{\langle{#1}}
\newcommand {\braket}[3]{\langle{#1}|{#2}|{#3}\rangle}
\newcommand {\bra}[1]{\langle{#1}|}

\newcommand{\e}{\!}

\begin{document}

\title{Overlapping resonances in the control of
intramolecular vibrational redistribution}

\author{D. Gerbasi$^1$}

\author{A. S. Sanz$^1$\footnote{Current address: Instituto de Matem\'aticas y F{\'\i}sica Fundamental,
Consejo Superior de Investigaciones Cient{\'\i}ficas, Serrano 123,
28006 Madrid, Spain}}

\author{P. S. Christopher$^1$}

\author{M. Shapiro$^2$}

\author{P. Brumer$^1$}
\affiliation{$^1$Chemical Physics Theory Group,
Department of Chemistry, \\
and Center for Quantum Information and Quantum Control, \\
University of Toronto, Toronto, Canada M5S 3H6;}

\affiliation{$^2$Chemical Physics Department,
The Weizmann Institute of Science,\\
Rehovot, Israel 76100, and Department of Chemistry,\\
University of British Columbia,
Vancouver, Canada, V6T 1Z1.}

\date{\today}

\begin{abstract}
Coherent control of bound state processes via the interfering
overlapping resonances scenario [Christopher {\it et al.},
J.\ Chem.\ Phys.\ {\bf 123}, 064313 (2006)] is developed to control
intramolecular
vibrational redistribution (IVR). The approach is applied to the
flow of population between bonds in a model of chaotic OCS
vibrational dynamics, showing the ability to significantly
alter the extent and rate of  IVR by
varying quantum interference contributions.
\end{abstract}

%\pacs{ }

\maketitle

%%%%%%%%%%%%%%%%%%%%%%%%%%%%%%%%%%%%%%%%%%%%%%%%%%%%%%%%%%%%%%%%%%%%%%%
%%%%%%%%%%%%%%%%%%%%%%%%%%%%%%%%%%%%%%%%%%%%%%%%%%%%%%%%%%%%%%%%%%%%%%%

\section{\label{sec1} Introduction}

Quantum control of molecular processes\cite{book1,book} has
proved, over the past two decades, to be viable both
theoretically and experimentally. An examination of the coherent
control literature, wherein scenarios are expressly designed to take
advantage of quantum interference phenomena, shows that the vast
majority of applications has been to processes occurring in the
 continuum energy regime.
Recently we proposed a new approach to
controlling bound state dynamics in large polyatomic molecules\cite{csb}
that exploits interferences between overlapping resonances.
We have demonstrated the
viability of this scenario in controlling internal conversion in
pyrazine.\cite{csb,csb2,csb3} In the present paper we further develop
this method, applying it to the
control of Intramolecular Vibrational Redistribution (IVR).
As an example, we study the control of the
flow of energy between bonds in a model of OCS. This molecule,
though small, is of particular interest at high energies, where,
classically, it displays predominantly chaotic
dynamics. In spite of the classical chaos, quantum control via
the present scenario
is shown to be excellent.

This paper is organized as follows: Section II provides an
overview of the theory, with a discussion of the Feshbach
partitioning technique which, as we have shown,\cite{csb}
provides a highly efficient
method for dealing with bound state problems. Section
III describes the collinear OCS model and its classical
dynamical characteristics. In Section IV we discuss the application of
the method to the control of IVR in OCS. An Appendix describes our use of
the Feshbach partitioning technique
for the numerical solution of the bound state problem for small systems
such as OCS.
A more ambitious
method for addressing considerably larger systems, the ``QP algorithm'',
is described elsewhere.\cite{csb2}

%%%%%%%%%%%%%%%%%%%%%%%%%%%%%%%%%%%%%%%%%%%%%%%%%%%%%%%%%%%%%%%%%%%%%%%
%%%%%%%%%%%%%%%%%%%%%%%%%%%%%%%%%%%%%%%%%%%%%%%%%%%%%%%%%%%%%%%%%%%%%%%

\section{\label{sec2}  Bound State Control}

%%%%%%%%%%%%%%%%%%%%%%%%%%%%%%%%%%%%%%%%%%%%%%%%%%%%%%%%%%%%%%%%%%%%%%%

\subsection{\label{sec2a} Time-evolution of populations in molecular
 systems}

We consider a system described by an Hamiltonian $H$ which can be partitioned
physically into the sum of two components $H_A$ and $H_B$, plus the
interaction $H_{AB}$ between them:
\begin{equation}
 H = H_A + H_B + H_{AB} ,
 \label{fullhamil}
\end{equation}
The eigenstates and eigenvalues of the full Hamiltonian
are defined by:
\begin{equation}
H \ket{\gamma}=  E_\gamma \ket{\gamma}~.
\end{equation}
The (``zeroth-order'') eigenstates and eigenvalues of the sum of the
decoupled Hamiltonians are defined as
\begin{equation}
(H_A + H_B) |\kappa\rangle = E^{(\kappa)} |\kappa\rangle~.
\end{equation}
Below, we are interested in the time evolution of the system, initially
prepared in a superposition of zeroth order states.
\beq
 |\Psi(t=0)\rangle = \sum_{\kappa} c_{\kappa} |\kappa\rangle ,
 \label{psi0}
\eeq
where $\{c_\kappa\}$ are ``preparation'' coefficients.
All sums over $\ket{\kappa}$, here
and below, are assumed to be confined to a subspace $S$. For example,
the selected initial states might consist of a set with
population heavily concentrated in one bond of a molecule, in which case,
energy flow out of such superposition states is examined.

The time-evolution of Eq. (\ref{psi0}) at any subsequent time can
then be obtained by expanding the (zeroth order) eigenstates,
$|\kappa\rangle$, in terms of the exact eigenstates $
\ket\gamma$ to give:
\beq
 \ket{\Psi(t)} = \sum_{\kappa,\gamma} c_{\kappa} a_{\kappa,\gamma}^*
  e^{-i E_{\gamma}t/\hbar}\ket{\gamma} ,
 \label{td}
\eeq
with $a_{\kappa,\gamma}^* = \brak{\gamma}\ket{\kappa}$.
The structure of $|\brak{\gamma}\ket{\kappa}|^2$ as
a function of $\gamma$ defines a resonance shape that
provides insight, in the
frequency domain, into the population flow out and into the zeroth order
$\ket{\kappa}$ states.

Given this time evolution,
the amplitude for finding the system in
a state $\ket{\kappa}$ at time $t$ is
\beq
 c_\kappa = \brak{\kappa}\ket{\Psi(t)} =
  \sum_{\kappa'} c_{\kappa'} M_{\kappa,\kappa'}(t) ,
 \label{time1}
\eeq
where
\begin{eqnarray}
 M_{\kappa,\kappa'}(t) & \equiv & \sum_{\gamma}
  a_{\kappa,\gamma}^{} a_{\kappa',\gamma}^* e^{-i E_{\gamma}t/\hbar}
  \nonumber \\
 & = & \brak{\kappa}| \left( \sum_{\gamma} e^{-i E_{\gamma}t/\hbar}
   \ket{\gamma} \brak{\gamma}| \right) \ket{\kappa'}
 \label{mkappa}
\end{eqnarray}
is the ($\kappa,\kappa'$) element of the overlap matrix
${\bf M}(t)$ defined by the term in brackets in Eq. (\ref{mkappa}).
Note that, for $\kappa' \neq \kappa$, if the states $\ket{\kappa}$
and $\ket{\kappa'}$ do not overlap with a common
$\ket{\gamma}$, i.e., there are no {\it overlapping
resonances}, then $M_{\kappa,\kappa'} = 0$. Our previous studies\cite{csb}
have demonstrated the significance of such overlapping resonances to the
control of radiationless transitions, such as internal conversion.

From Eq.~(\ref{time1}), the probability of finding the system in a
collection of states $\ket{\kappa}$ contained in the initial set $S$
at time $t$ is given by
\begin{equation}
 P(t) = \sum_{\kappa} |\brak{\kappa}\ket{\Psi(t)}|^2 =
  {\bf c}^{\dagger} {\bf K}(t) {\bf c} ,
 \label{totalpop2}
\end{equation}
where ${\bf c}$ is a $\kappa$-dimensional vector whose components
are the $c_{\kappa}$ coefficients, and ${\bf K}(t) \equiv {\bf
M}^\dagger (t){\bf M}(t)$. The generalization to the question of
finding population in an alternative collection of states, other
than $S$, is straightforward. However, it is unnecessary for the
study below, as will become evident.  Equation~(\ref{totalpop2})
allows us to address the question of enhancing or restricting the
flow of probability out of $S$ by finding the optimal combination
of $c_{\kappa}$ that achieves this goal at a specified time $T$.
Experimentally, the resultant required superposition state can be
prepared using modern pulse shaping techniques.

%%%%%%%%%%%%%%%%%%%%%%%%%%%%%%%%%%%%%%%%%%%%%%%%%%%%%%%%%%%%%%%%%%%%%%%

\subsection{\label{sec2b} The Feshbach partitioning technique}

Our interest is to control the flow of
population out of some generic molecular subspace into the
entire molecular Hilbert space.
In order to do so we make use of the bound state version of the
Feshbach partitioning technique.\cite{f,Levine}
Here, since the control approach
is being tested on a small system, we solve the resulting equations in
a straightforward way, as described in Appendix A. Larger systems can
take advantage of the ``QP algorithm''.\cite{csb2}

The Feshbach partitioning  technique is based on defining two
projection operators
\beq
 Q \equiv \sum_{\kappa} \ket{\kappa}\bra{\kappa} , \quad
 P \equiv \sum_{\beta} \ket{\beta} \bra{\beta} ,
 \label{qpoperators}
\eeq
which satisfy the following properties:
\begin{subequations}
 \begin{eqnarray}
  & Q^2 = Q , \quad P^2 = P , &
  \label{equal1} \\
  & [Q,P] = 0 , &
  \label{equal2} \\
  & P + Q = \mathbb{I} , &
  \label{equal3}
 \end{eqnarray}
 \label{equal}
\end{subequations}
where $\mathbb{I}$ is the identity operator.
In what follows, the flow of probability of interest is from the
$Q$ space to the $P$ space.

Using Eqs. (\ref{equal3}) and (\ref{qpoperators}), the eigenstates of
the full Hamiltonian can be written as
\beq
 \ket{\gamma} = \sum_{\kappa} \ket{\kappa}\brak{\kappa}\ket{\gamma}
  + \sum_{\beta} \ket{\beta}\brak{\beta}\ket{\gamma}.
\eeq
Similarly, the Schr\"odinger equation can be expressed as
\beq
 [E_{\gamma} - H][P + Q]\ket{\gamma} = 0 ,
\eeq
whereby multiplying it by $P$ and then by $Q$, and using Eq.
(\ref{equal}), one obtains the following set of coupled equations:
\begin{subequations}
 \begin{eqnarray}
  \left[E_{\gamma} - PHP\right]P\ket{\gamma} & = & PHQ\ket{\gamma} ,
  \label{p1} \\
  \left[E_{\gamma} - QHQ\right]Q\ket{\gamma} & = & QHP\ket{\gamma} .
  \label{q1}
 \end{eqnarray}
 \label{coupledset}
\end{subequations}

The states $\ket{\kappa}$ and $\ket{\beta}$ are solutions to the
decoupled (homogeneous) equations arising from Eqs.~(\ref{p1}) and
(\ref{q1}), respectively.
That is,
\begin{subequations}
 \begin{eqnarray}
  \left[E_{\beta} - PHP\right]P\ket{\beta} = 0 ,
  \label{homobeta} \\
  \left[E_{\kappa} - QHQ\right]Q\ket{\kappa} = 0.
  \label{homokappa}
 \end{eqnarray}
 \label{homos}
\end{subequations}
Contrary to continuum problems, in general $E_{\gamma} \ne E_{\beta}$ and
it is possible to express $P\ket{\gamma}$ in terms of
the particular solution of the (inhomogeneous) Eq. (\ref{p1}),
\begin{equation}
 P\ket{\gamma}  = [E_{\gamma} - PHP]^{-1}PHQ\ket{\gamma} .
 \label{psolve}
\end{equation}

Substituting Eq.~(\ref{psolve}) into Eq.~(\ref{q1}) results in
\beq
 [E_{\gamma} - QHQ]Q\ket{\psi} =
  QHP[E_\gamma - PHP]^{-1}PHQ\ket{\psi} .
\eeq
By rearranging terms in this equation, one obtains
\beq
 [E_{\gamma} - {\cal H}]Q\ket{\gamma} = 0 ,
 \label{qket}
\eeq
where ${\cal H}$ is an effective Hamiltonian, defined as
\beq
 {\cal H} = QHQ + QHP[E_{\gamma} - PHP]^{-1}PHQ .
\label{hbar}
\eeq
The term between squared brackets can be written as
\beq
 \left[E_{\gamma} - PHP\right]^{-1} = \sum_{\beta}
  \frac{1}{E_{\gamma} - E_{\beta}} \ket{\beta}\bra{\beta}
 \label{specres}
\eeq
by using the spectral resolution of an operator.
The matrix elements of ${\cal H}$ are given by
\beq
 \braket{\kappa}{{\cal H}}{\kappa'} =
  E_{\kappa} \delta_{\kappa, \kappa'} + \Delta_{\kappa,\kappa'} ,
\eeq
where
\begin{subequations}
 \begin{eqnarray}
  \Delta_{\kappa,\kappa'} = \frac{1}{2\pi} \sum_{\beta}
   \frac{\Gamma_{\kappa,\kappa'}}{{E_{\gamma} - E_{\beta}}} ,
  \label{delta} \\
  \Gamma_{\kappa,\kappa'} = 2 \pi V(\kappa|\beta) V(\beta|\kappa') ,
  \label{gamma}
 \end{eqnarray}
 \label{parameters}
\end{subequations}
with $V(\kappa|\beta) = \braket{\kappa}{QHP}{\beta}$ being the
coupling term.  Equations~(\ref{delta}) and (\ref{gamma})
represent the energy shift and the decay rate, respectively. By
diagonalizing Eq.~(\ref{qket}) in a self-consistent manner, one
obtains the energy eigenvalues, $E_{\gamma}$, and the values for
the overlap integrals, $a_{\kappa, \gamma}$.

\begin{table}[t]
 \begin{tabular}{c p{.3cm} c p{.3cm} c p{.3cm} c}
  \hline\hline
   \ $i$ & & $D_i$   & & $\beta_i$ & & $R_i^0$ \\ \hline
   \ 1   & & 0.08518 & & 1.5000    & & 2.9759 \\
   \ 2   & & 0.21238 & & 1.6251    & & 2.2559 \\
   \ 3   & & 0.16000 & & 1.1589    & & 2.8037 \\
  \hline\hline
 \end{tabular}
 \caption{Parameters defining the potential energy surface given by
  Eq.~(\ref{eq04}).
  All magnitudes are given in a.u.}
 \label{tab1}
\end{table}

Note that the energy eigenvalues and the overlap integrals can also be
obtained\cite{thesis} by directly diagonalizing the full Schr\"odinger
equation in the zeroth-order basis.
However, the partitioning technique presented above has computational
advantages for cases where the dimensions of the $P$ space is large,
since one only needs to diagonalize an effective Hamiltonian ${\cal H}$
with dimensions given by the $Q$ space.
Note, however, that diagonalizing ${\cal H}$ requires using iterative
procedures, and needs to be solved repeatedly until each eigenvalue is
found. Appendix~\ref{secapA} provides details on the partitioning
algorithm used here.

%%%%%%%%%%%%%%%%%%%%%%%%%%%%%%%%%%%%%%%%%%%%%%%%%%%%%%%%%%%%%%%%%%%%%%%

\subsection{\label{sec2c} Optimal Control}

To  determine the set of optimal preparation coefficients
leading to either a maximum or a minimum population at time $t = T$,
 we need to find the extrema of the function\cite{fs}
\beq
 P_\lambda (t) = {\bf c}^\dagger {\bf K}(t) {\bf c}
  - \lambda {\bf c}^\dagger {\bf c}
 \label{maximize}
\eeq
with respect to the coefficients ${\bf c}$, where $\lambda$ is a
Lagrange multiplier added to assure normalization, i.e.,
\beq
 \sum_\kappa |c_\kappa|^2 = 1 .
\eeq
The optimum vector, ${\bf c}_T$, is obtained by differentiating
Eq.~(\ref{maximize}) with respect to the components of
${\bf c}^\dagger$, and equating the resulting expression to zero at
time $T$, i.e.,
\beq
 \frac{\partial P_\lambda (t)}{\partial c_{\kappa'}^*} \bigg|_{t = T}
  = \sum_\kappa {\bf K}_{\kappa',\kappa}(T)c_{\kappa}
   - \lambda c_{\kappa} = 0 .
 \label{optvec}
\eeq
The optimum vector resulting from this procedure is a solution to
the eigenvalue problem represented by Eq.~(\ref{optvec}). Note
that this vector can either maximize or minimize the solution. In
the first case, the interference between overlapping resonances
created by the initial superposition will be seen to result in a
delay of the population decay, whereas in the second case the population
decay is being accelerated.

In order to further clarify the dependence of the time-evolution on
overlapping resonances, Eq.~(\ref{totalpop2}) can be re-expressed as

\beq
 P(t) = \sum_\kappa |c_\kappa|^2 g_\kappa +
 \sum_{\substack{\kappa', \kappa \\ \kappa' \neq \kappa}}
  c_{\kappa'}^*c_{\kappa} f_{\kappa',\kappa} ,
 \label{totalpop3}
\eeq
where
\beq
 g_\kappa = \sum_{\kappa'} |M_{\kappa',\kappa}|^2 ,
 \label{gterm}
\eeq
and
\beq
 f_{\kappa',\kappa} = \sum_{\kappa''} M^*_{\kappa'',\kappa'}
  M_{\kappa'',\kappa} .
\label{fterm}
\eeq
As expressed in Eq.~(\ref{totalpop3}), the $Q$ space population
assumes the generic coherent control form\cite{book,book1}:
it is given as the sum of non-interfering pathways,
represented by $g_{\kappa'}$, and interfering pathways,
represented by $f_{\kappa',\kappa}$.

%%%%%%%%%%%%%%%%%%%%%%%%%%%%%%%%%%%%%%%%%%%%%%%%%%%%%%%%%%%%%%%%%%%%%%%

\subsection{The Role of Overlapping Resonances}

The interference term in Eq.~(\ref{totalpop3}) depends on
$f_{\kappa',\kappa}$, which, in accord with Eq. (\ref{fterm}),
depends upon the overlap between resonances. Qualitatively
speaking,  a resonance is described by bound states
$\ket{\kappa}$ coupled to a quasi-continuum of exact eigenstates
$\ket{\gamma}$. Each such state is thus associated with the
energy width of the $\langle\kappa|\gamma\rangle$ overlap coefficients.
Overlapping resonances are the result of having at
least two states whose resonance widths are wider than their
associated level spacing. The resulting resonances interfere with
one another, displaying a variety of lineshapes,\cite{book} and
are responsible for the interference in this control scenario. In
the absence of overlapping resonances the full
$f_{\kappa',\kappa}$-term in Eq.~(\ref{totalpop3}) vanishes and
control disappears.

Note that there are also
contributions from overlapping resonances to the $g_\kappa$-term, as
can be seen from their effect on the nature of the decay from the
individual $\ket{\kappa}$.
These resonances distort the lineshape, and hence
the corresponding time dependence.
In order to determine the contribution from overlapping resonances, we
have devised\cite{csb} a qualitative measure, defined as
\beq
 \tilde{P}(t) = [P(t) - W(t)] ,
 \label{poverlap}
\eeq
where
\beq
 W(t) = \sum_{\kappa} |c_\kappa M_{\kappa,\kappa}|^2 .
\label{wterm}
\eeq
Here $W(t)$ is a measure of the direct contribution, and $\tilde{P}(t)$
provides a measure of the overlapping resonance contribution.
In the absence of overlapping resonances, $P(t) = W(t)$.

%%%%%%%%%%%%%%%%%%%%%%%%%%%%%%%%%%%%%%%%%%%%%%%%%%%%%%%%%%%%%%%%%%%%%%%
%%%%%%%%%%%%%%%%%%%%%%%%%%%%%%%%%%%%%%%%%%%%%%%%%%%%%%%%%%%%%%%%%%%%%%%

\section{\label{sec3} Classical aspects of the collinear OCS}

%%%%%%%%%%%%%%%%%%%%%%%%%%%%%%%%%%%%%%%%%%%%%%%%%%%%%%%%%%%%%%%%%%%%%%%

\subsection{\label{sec3a} The OCS model}

As a working model to illustrate the usefulness of the method described
in Sec.~\ref{sec2}, we consider a collinear model of OCS, with a modified
Sorbie-Murrell\cite{cb}  potential.
The interest in this system arises from the fact that, close to
dissociation, i.e. in the energy region of interest below,
the classical dynamics becomes highly chaotic.
As such, collinear OCS is a complex system with a penchant for
extensive IVR.

The classical dynamics of  OCS has been
studied in both planar,\cite{cb} and collinear
\cite{Davis1,Davis2} versions.
Here, we consider the collinear case, which is described by the
Hamiltonian
\begin{equation}
 H = \frac{P_1^2}{2 \mu_{13}} + \frac{P_2^2}{2 \mu_{23}} -
  \frac{P_1 P_2}{m_C} + V (R_1, R_2, R_3) ,
 \label{eq03}
\end{equation}
where
\begin{eqnarray}
 \mu_{13} & = & \frac{m_O m_C}{m_O + m_C} \nonumber \\
  \mu_{23} & = & \frac{m_S m_C}{m_S + m_C} ,
\end{eqnarray}
are reduced masses; $R_1$ and $R_2$ are the CS and CO bond
distances, respectively ($R_3 = R_1 + R_2$); and $P_1$ and $P_2$
are  the corresponding momenta.

\begin{figure}[t]
 \includegraphics[width=5.5cm]{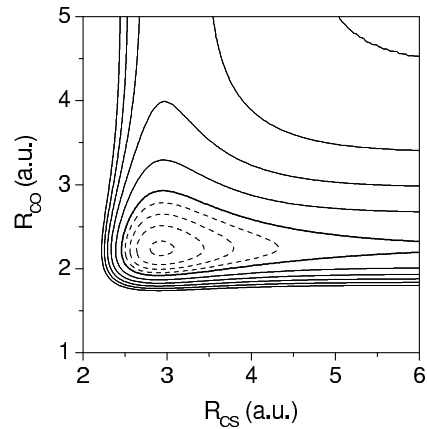}
 \caption{Contour plot of the potential energy surface given by Eq.\
  (\ref{eq04}).
  Solid and dashed lines represent, respectively, energy contours above
  and below the dissociation onset, at $E = 0.100$ a.u.\ (thick solid
  line).}
 \label{fig1}
\end{figure}

In the course of this work we found that the
Sorbie-Murrell
OCS model\cite{cb} displayed a second minimum at large
distances along both the CS and CO exit channels.
Although the depth of this second well is extremely small, there
are a large number of closely packed eigenstates localized in
this region due to the length of the well.
To our knowledge, there is no experimental evidence
to either support or refute a second minimum, although they have
been associated\cite{ozone} with van der Waals interactions  in
O$_3$. However, in order to ensure that the observed control is
not a manifestation of this secondary minimum (as was the case in
our preliminary studies) a modified interaction potential is used
that removes these second minima while retaining the general
features of the remaining potential. Specifically, the potential
used  here consists of a sum of three Morse functions,
\begin{equation}
 V(R_1, R_2, R_3) \equiv \sum_{i = 1}^3 V_i =
  \sum_{i = 1}^3 D_i \left[ 1 - e^{- \beta_i (R_i - R_i^0)} \right]^2 ,
 \label{eq04}
\end{equation}
with parameters given in Table \ref{tab1}. A contour plot of the
resultant potential energy surface is shown in Fig.~\ref{fig1}.
Except for the Morse function $V_3$, which depends on $R_3$,
the parameters defining
the other two Morse functions have been changed so that the potential smoothly
fits the original  one along the equilibrium directions while,
at the same time, eliminating the second potential minima.
Moreover, we have also modified the  added  constant in the potential
so that the CS dissociation onset [$V(\infty, R_2^0) =
D_1 + D_3$] corresponds to the original value of $E_d = 0.100$ a.u.

%%%%%%%%%%%%%%%%%%%%%%%%%%%%%%%%%%%%%%%%%%%%%%%%%%%%%%%%%%%%%%%%%%%%%%%

\subsection{\label{sec3b} Characterizing the Dynamics}

\begin{figure}
 \begin{center}
 \includegraphics[width=5.5cm]{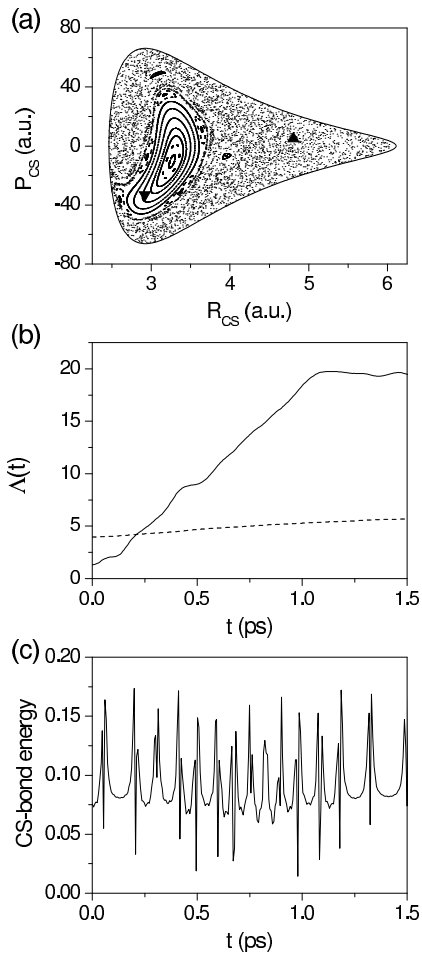}
 \caption{\label{fig2} (a) Poincar\'e surface of section for the
  collinear OCS model at $E = 0.09796$ a.u.
  The solid line represents the total energy contour.
  (b) Distance between two nearby trajectories (with $d_0 = 10^{-8}$
  a.u.) chosen in the stable region ($\blacktriangledown$), and in the
  chaotic sea ($\blacktriangle$).
  The high-frequency oscillations have been averaged out in both cases
  (the smoothing causes $\Lambda^{(\blacktriangledown)}$ to appear as
  if it does not begin at zero).
  (c) CS--bond vibrational energy corresponding to the chaotic
  trajectory of \mbox{part (b)}.}
 \end{center}
\end{figure}

The classical dynamics of the resultant OCS model is
characterized by a smooth transition from regular to chaotic
dynamics with increasing energy. At an energy just below
dissociation, ($E_0 = 0.09796$4~a.u., of interest below), the
Poincare surface of section\cite{PSOS} shows [Fig.~\ref{fig2}(a)]
highly chaotic dynamics, with a stable region constituting about
$1/3$ of the phase-space portrait. This energy corresponds to the
mean value of the energies of the two wave packets,
$|\Psi_{\pm}\rangle$, obtained below in maximizing and minimizing
the energy flow from the CS  bond. Surfaces of section in the
nearby energies are essentially similar. This being the case,
there is no obvious classical origin to the control of bond
energy relaxation described below. Of some future interest,
however, might be an auxiliary study of the relationship of
overlapping resonances induced control, observed below, to
classical features such as bond energy recurrences, cantori, and
the inhomogeneous character  of the OCS phase space
\cite{Eckhardt1,Davis1,Davis2,Chirikov,Wyatt,Davis3}.

Quantitative insight into the rate of loss of correlations in the
chaotic region of phase space can be obtained by computing Lyapunov
exponents,\cite{Lieberman} approximated by the average (over various
trajectories) of the
exponential rate at which the distance $d(t)$ between adjacent
trajectories in phase space grow in time:
\begin{equation}
 \lambda_\infty = \lim_{t \to \infty}
  \frac{1}{t} \ln \frac{d(t)}{d_0} .
 \label{eq:1}
\end{equation}
Here, in order to show how different regular and chaotic trajectories
behave, we have computed the quantity
\begin{equation}
 \Lambda (t) = \ln \frac{d(t)}{d_0} ,
 \label{short}
\end{equation}
with $d_0 = 10^{-8}$~a.u.
We label the finite time
Lyapunov exponent, computed in this way, as $\lambda_t$.

The quantity $\Lambda (t)$ is shown in Fig.~\ref{fig2}(b) for two
sets of nearby trajectories,\cite{note2} picked in two
different regions of phase-space: the stable
island, and the chaotic sea. The results, to
$t \approx 1.2$~ps, give
$\lambda_t^{(stable)} \simeq 1.46$~ps$^{-1}$, and
$\lambda_t^{(chaotic)} \simeq 17.41$~ps$^{-1}$ in the
regular
and chaotic regions, respectively. The associated times are to
be compared to zeroth order vibrational periods (27.45 fs
for the CS bond, and 18.10 fs for the CO bond).

Finally, in Fig.~\ref{fig2}(c) we show the energy in the CS bond,
for a trajectory in the chaotic sea.
As can be seen, the energy displays a complicated pattern, with
irregular energy transfer between both
bonds as a function of  time.
Nonetheless, when one computes the energy average of an ensemble of
trajectories, the pattern becomes smooth and displaying a profile than
can be fitted to an exponential decay,\cite{Davis2} similar to
those observed in its quantum counterpart below.

%%%%%%%%%%%%%%%%%%%%%%%%%%%%%%%%%%%%%%%%%%%%%%%%%%%%%%%%%%%%%%%%%%%%%%%
%%%%%%%%%%%%%%%%%%%%%%%%%%%%%%%%%%%%%%%%%%%%%%%%%%%%%%%%%%%%%%%%%%%%%%%

\section{\label{sec4} Coherent control of IVR}

%%%%%%%%%%%%%%%%%%%%%%%%%%%%%%%%%%%%%%%%%%%%%%%%%%%%%%%%%%%%%%%%%%%%%%%

\subsection{\label{sec4a} Population decay}

\begin{table*}
 \begin{ruledtabular}
 \begin{tabular}{c p{0.3cm} c p{0.3cm} c c p{0.25cm} c c p{0.5cm} c p{0.3cm} c c p{0.25cm} c c p{0.5cm} c p{0.3cm} c}
  \multicolumn{4}{c}{} & \multicolumn{7}{c}{IVR suppression} & & \multicolumn{7}{c}{IVR enhancement} & & \\
  \cline{5-11} \cline{13-19}
  \ $\kappa$&&$E_{\kappa}$ (a.u.) &&& $c_\kappa^r$ &&& $c_\kappa^i$ && $|c_\kappa|^2$ &&&
                          $c_\kappa^r$ &&& $c_\kappa^i$ && $|c_\kappa|^2$ && $t_\delta$~(fs) \\
  \hline
   \ 1 &&0.0851446&&      &0.02895&&      &0.00000 && 0.00084 && $-$\e& 0.13839 &&      & 0.00000 &&  0.01915 && 17.84 \\
   \ 2 &&0.0850268&& $-$\e&0.00706&&      &0.17289 && 0.02994 &&      & 0.38027 && $-$\e& 0.00806 &&  0.14467 && \hspace{.055cm} 8.03 \\
   \ 3 &&0.0848265&&      &0.16188&& $-$\e&0.16611 && 0.05380 && $-$\e& 0.00128 && $-$\e& 0.10472 &&  0.01097 && 16.25 \\
   \ 4 &&0.0845437&& $-$\e&0.56608&&      &0.25828 && 0.38716 &&      & 0.01257 && $-$\e& 0.08349 &&  0.00713 && 20.76 \\
   \ 5 &&0.0841783&&      &0.18017&& $-$\e&0.24251 && 0.09127 && $-$\e& 0.03560 &&      & 0.05674 &&  0.00449 && 13.69 \\
   \ 6 &&0.0837303&&      &0.20267&&      &0.15178 && 0.06411 &&      & 0.19804 &&      & 0.05108 &&  0.04183 && 34.02 \\
   \ 7 &&0.0831998&& $-$\e&0.21171&&      &0.16534 && 0.07216 &&      & 0.12120 && $-$\e& 0.63185 &&  0.41392 && 20.53 \\
   \ 8 &&0.0825867&&      &0.25004&& $-$\e&0.41477 && 0.23455 &&      & 0.20859 && $-$\e& 0.39177 &&  0.19699 && 25.32 \\
   \ 9 &&0.0818910&& $-$\e&0.05482&&      &0.25131 && 0.06616 &&      & 0.24895 && $-$\e& 0.31440 &&  0.16082 && 22.95 \\
 \end{tabular}
 \end{ruledtabular}
 \caption{Values corresponding to the eigenstates for the (uncoupled)
  CS bond used in the optimized superpositions.
  $E_\kappa$ denotes the energy associated to these eigenstates;
  $c_\kappa^r$ and $c_\kappa^i$ are the real and imaginary parts of
  the $c_\kappa$ coefficients, respectively; and $t_\delta$ is the
  decay time (see text for details).
  The optimization to maximize/minimize the energy transfer into the CO
  bond (suppression/enhancement of IVR) has been carried out at
  $T = 100$~fs.
  The energy corresponding to the ground state in the (uncoupled)
  CO bond is $E_{CO}^0 = 0.00360475$~a.u.}
 \label{tab2}
\end{table*}

We now consider the suppression (and enhancement) of IVR in the above
model of OCS. Our intent is
to assess the extent of control in such a system, and to establish the
relationship between control and
overlapping resonances.
The coupling terms $V(\kappa|\beta)$ and, subsequently, the overlap
integrals $a_{\kappa,\gamma}$ and the energy eigenvalues $E_{\gamma}$ are
calculated by expanding the OCS wave functions in products of
the zeroth order states,
\beq
 \ket{\Psi} = \sum_{m,n} \ket{\eta_{CS}^m} \otimes \ket{\xi_{CO}^n}d_{mn}~.
 \label{nocoupling}
\eeq
where $\ket{\eta_{CS}^m}$ and $\ket{\xi_{CO}^n}$ are eigenstates of the
uncoupled CS and CO bond potentials, respectively, with quantum numbers
$m$ and $n$.
Our interest is in the flow, for example, out of the CS bond. Hence,
the $Q$ subspace is chosen to represent all wave functions
containing only excitation in the CS bond, i.e., $\ket{\kappa}$
are
$ \ket{\eta_{CS}^m} \otimes \ket{\xi_{CO}^0}$,
for all $m$, whereas the $P$ subspace  spans the space
represented by all other zeroth order excitations, i.e., the
$\ket{\beta}$ are
$ \ket{\eta_{CS}^m} \otimes \ket{\xi_{CO}^n}, n \neq 0$,
 describing excitation in the
CS bond. Initiating excitation within $Q$ and watching the flow
into $P$ then corresponds to an experiment wherein excitation
flows out of the CS bond.

As seen in Sec.~\ref{sec3a}, the coupling term, $QHP$, necessary
to obtain the energy shifts and decay rates, consists of a static
term ($V_{3}$), and a dynamic term [proportional to $p_1 p_2$ in
Eq. (\ref{eq03})]. The overlap integrals and energy eigenvalues
are obtained by self-consistent diagonalization of
Eq.~(\ref{qket}). All vibrational states, $\ket{\eta_{CS}^m}$ and
$\ket{\xi_{CO}^n}$, are numerically calculated using a discrete
variable representation (DVR) technique,\cite{Light} obtaining a
total of 45 eigenvectors for the CS bond, and 59 for the CO bond.
The number of eigenvectors is larger in the second case, because
the dissociation threshold of the CO bond is higher in energy.

From all the vibrational states obtained, we have observed that
control is best when considering a superposition of states, i.e.,
Eq. (\ref{psi0}), that is near the dissociation onset. The energy
differences between these states are relatively small ($\approx
0.0004$ a.u., whose inverse corresponds to a timescale of
$\approx 60$ fs), thus giving rise to a high density of states
with time scales comparable to vibrational relaxation. The result
is a greater opportunity for overlapping resonances which, as
will be seen below, enhances the ability to control energy flow.
In our case, the states used are the last nine bound
eigenvectors (before the dissociation onset) of the
CS bond, whose corresponding eigenvalues are given in
Table~\ref{tab2}. Note, however, that dense eigenstate manifolds
will occur at far lower energies in larger molecules.  Hence, the
initial $\ket{\Phi(0)}$ is comprised of a superposition of nine
CS states in Table 1, with the CO in the ground vibrational state.

\begin{figure}[b]
 \includegraphics[width=10cm,angle=90]{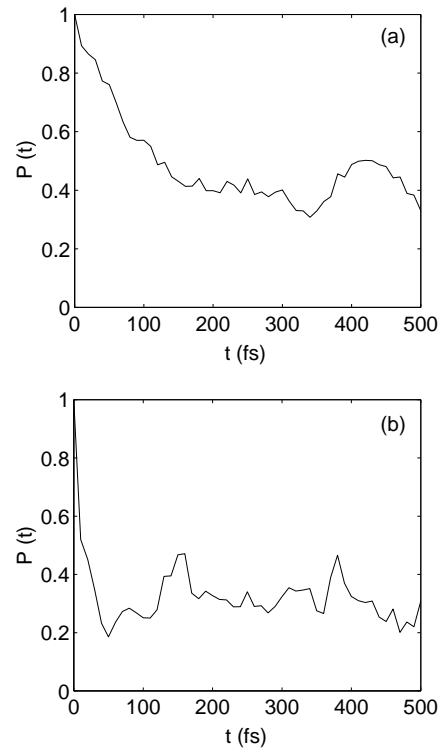}
 \caption{IVR control in OCS: (a) IVR suppression, and (b) IVR
  enhancement.
  The parameters defining the optimal superposition for $T = 100$~fs
  are given in Table~\ref{tab2}.}
 \label{fig3}
\end{figure}

\begin{figure*}
 \includegraphics[width=12cm,angle=90]{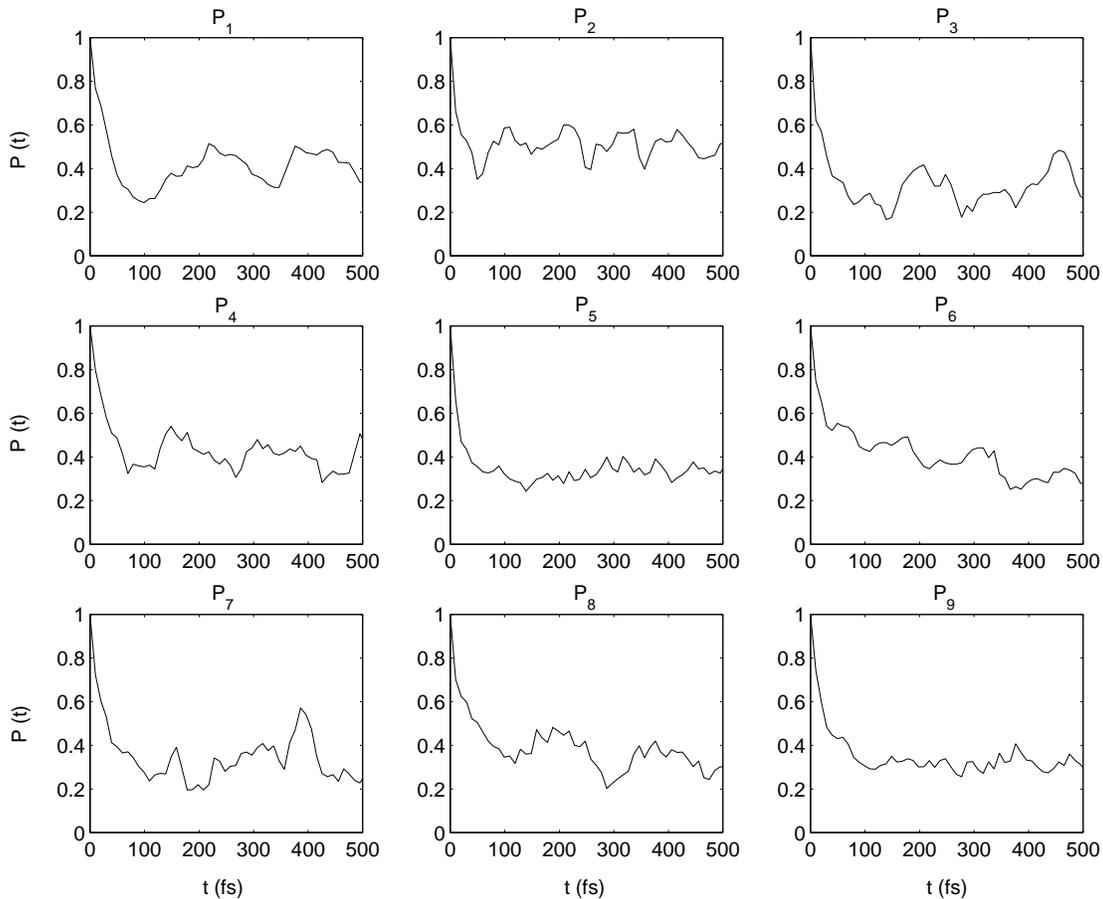}
 \caption{Individual decay for wave functions consisting of each
  individual eigenvector used in the construction of the optimal
  superpositions.
  The labels correspond to those given in Table~\ref{tab2}.}
 \label{fig4}
\end{figure*}

Figure~\ref{fig3}  shows the
time-evolution of the population, $P(t)$, for an initial
wave function constructed from the nine zeroth order $Q$ space
states noted above, and optimized for maximal or minimal energy
flow at $T=100$ fs. The optimal coefficients were found using the
method described in Sec.~\ref{sec2}; the $c_\kappa$ coefficients
and their probabilities are given in Table~\ref{tab2}. Results in
panel (a) correspond to an
 initial superposition optimized to minimize
the population flow from the $Q$ to the $P$ space, while panel (b)
shows results optimized to enhance the flow of population.
As is clearly seen, the initial falloff in panel (a) is much slower
than that in (b).
To quantify this decay, the initial  $P(t)$ falloff was fit
to an exponentially decreasing function,
\begin{equation}
 P(t) = P_\infty + (1 - P_\infty) e^{-t/t_\delta} ,
 \label{relax}
\end{equation}
where $t_\delta$ is the decay time, and $P_\infty$ is the average
around which $P(t)$ fluctuates for the first 1.0~ps.
Note that the $t_\delta$ values can only be regarded as approximate
since the falloff is, in general, not exponential, and $t_\delta$
depends on the time scale over which the exponential is fit.
(Here the fit is over 400 fs).
In case (a),
the decay time is $t_\delta \simeq 57.35$~fs, while in case (b)
it is $t_\delta \simeq 8.60$~fs, about seven times smaller.
Furthermore, we note that in panel (a), only about $24$\% of the
population has been transferred from $Q$ to $P$ during the first
50~fs, while, in contrast, approximately 82\% of the population
has being transferred to the $P$ in panel (b) during the same
time. Moreover, the population that asymptotically remains
localized along the CS bond is also larger in the case of IVR
suppression ($P_\infty \simeq 0.4$) than in that of enhancement
($P_\infty \simeq 0.3$).

The controlled results should be compared to the natural IVR
behavior of the individual levels participating in the
superposition. To this end, the $P(t)$ for each of the
participating levels is shown in Fig.~\ref{fig4}.
Although the
energy difference between these states shown is relatively small,
the populations, $P_\kappa$, evolve with a range of initial
falloff values, as can be seen in the corresponding values of
$t_\delta$, given in Table~\ref{tab2}.
Note also, from this table, that the control seen in
Fig.~\ref{fig3} is not due to the identification  of a particular
$|\kappa\rangle$ that independently maximizes or minimizes the
decay. Indeed, by inspecting the value of the $c_\kappa$
coefficients, we find, in the case of IVR suppression,
participation of most of the nine levels, with $\approx$  60\% of
the total initial population  concentrated in the two states with
$\kappa = 4$ and $\kappa = 8$. Neither of these two states
independently have the longest decay times, but their
interference is crucial to control. Similar observations result
from considering the data for optimized IVR enhancement, despite
the fact that $\kappa=2$ has a relatively small $t_\delta$. In this
case the optimized superposition also gives a significantly smaller
$P(T)$ than does the individual $\kappa=2$ state.

\begin{figure}[t]
 \includegraphics[width=10cm,angle=90]{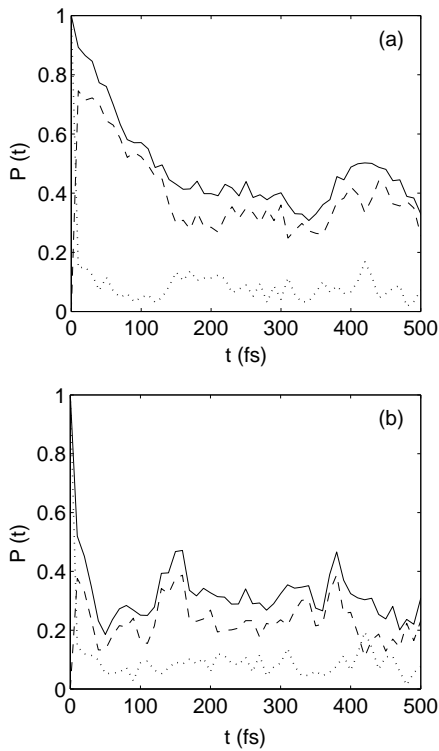}
 \caption{Contribution of $\tilde{P}(t)$ (dashed line) and $W(t)$
  (dotted line) to: (a) IVR suppression, and (b) IVR enhancement.
  The solid line represents the corresponding $P(t)$ function from
  Figs.~\ref{fig3}.}
 \label{fig5}
\end{figure}

A qualitative measure $\tilde{P}(t)$ of the contribution from the
interference of overlapping resonances, and $W(t)$ from the
direct contribution,  was provided in Eq.~(\ref{poverlap}).
Results for $\tilde{P}(t)$ and $W(t)$ for  the maximization
and minimization cases above are provided in Fig.~\ref{fig5}
where the contribution from overlapping resonances (dashed line),
become dominant after the first 10~fs, thus demonstrating the
important role played by these resonances in the IVR control
scenario. This is seen to be the case for both the maximization,
as well as minimization, of the flow from the CS bond.

\begin{figure}
 \begin{center}
 \includegraphics[width=8.5cm]{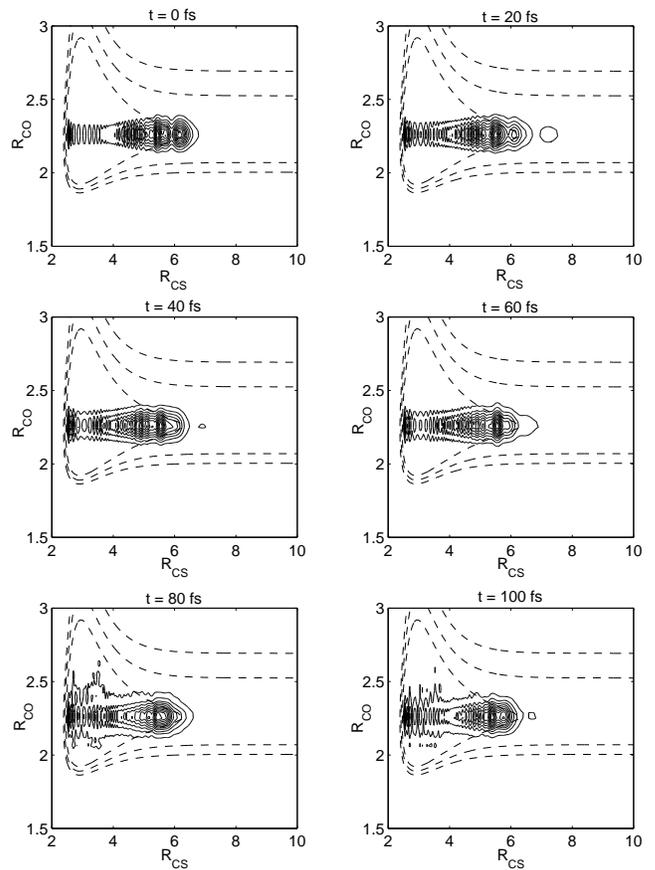}
 \caption{\label{fig7} Wave packet evolution corresponding to IVR
  suppression.
  Dashed lines represent equipotential energy contours, with the
  innermost corresponding to the wave packet energy,
  $E_{+} = 0.09849$~a.u.}
 \end{center}
\end{figure}

\begin{figure}
 \includegraphics[width=8.5cm]{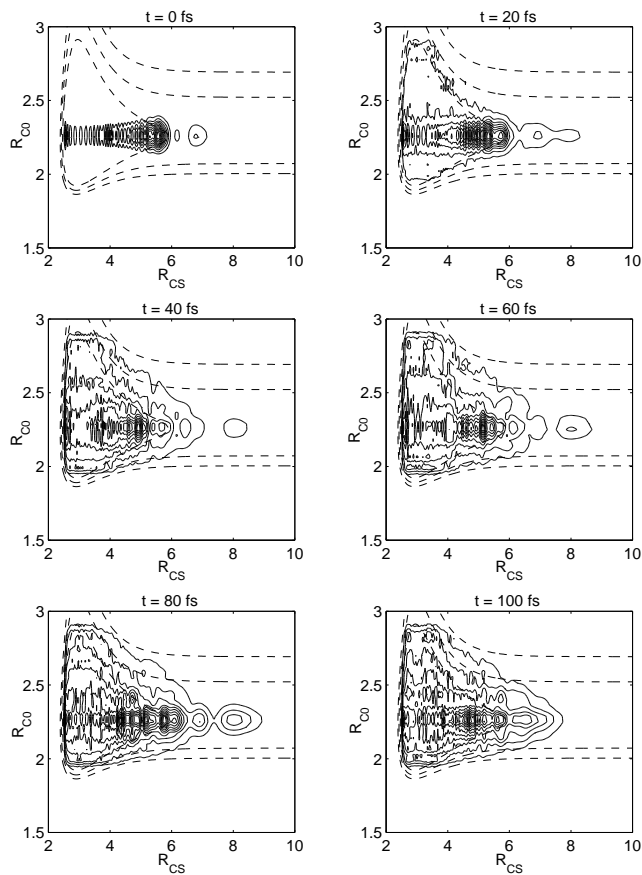}
 \caption{Wave packet evolution corresponding to IVR enhancement.
  Dashed lines represent equipotential energy contours, with the
  innermost corresponding to the wave packet energy,
  $E_{-} = 0.09743$~a.u.}
 \label{fig8}
\end{figure}

A pictorial, and enlightening, view of the results is provided in
Figs.~\ref{fig7} and \ref{fig8}, where the wave packets associated
with IVR suppression and enhancement are shown. As can be seen in
Fig.~\ref{fig7}, for the case of IVR suppression, the wave packet
remains highly localized along the $R_{\rm CS}$ mode, with
minimum spreading along the $R_{\rm CO}$ mode. In particular, it
undergoes a slight oscillation along the $R_{\rm CS}$ mode,
concentrating most of the probability around the region where the
CS dissociation takes place, in a clear correspondence to what
happens with a classical counterpart. For the case of IVR
enhancement, the effect is the opposite. As can be seen in
Fig.~\ref{fig8}, the spreading of the wave packet along the
$R_{\rm CO}$ mode coordinate is relatively fast.

The method described above is, of course, applicable at any time
during the dynamics. For example, we tried, and successfully
attained, control for times at long as 1.5 ps (corresponding to
over 50 CS vibrational periods), resulting in about a 55\% of
the population localized in the CS bond for IVR suppression, and
about 22\% for IVR enhancement.

%%%%%%%%%%%%%%%%%%%%%%%%%%%%%%%%%%%%%%%%%%%%%%%%%%%%%%%%%%%%%%%%%%%%%%%
%%%%%%%%%%%%%%%%%%%%%%%%%%%%%%%%%%%%%%%%%%%%%%%%%%%%%%%%%%%%%%%%%%%%%%%

\section{\label{sec5} Comments and Summary}

In this paper,
a method for controlling intramolecular vibrational
redistribution has been developed and has been applied to
OCS, where extensive control over IVR is
attained. Of particular interest is that the control is
achieved even though the associated classical dynamics is
chaotic.
The method, wherein the coefficients of an initial
superposition of zeroth order states are optimized, is shown to
rely upon the presence of overlapping resonances, a feature which
is expected to be ubiquitous in realistic molecular systems.

We have assumed throughout this paper that the initial state
that optimizes the intramolecular vibrational
redistribution can be prepared, for a real molecule,
using modern pulse shaping techniques. Computations displaying
the resultant field were not, however, carried out on this OCS model
since they are best done using more realistic molecular potentials
in higher dimensions, yielding realistic optimizing fields.
Work of this kind is in progress.

%%%%%%%%%%%%%%%%%%%%%%%%%%%%%%%%%%%%%%%%%%%%%%%%%%%%%%%%%%%%%%%%%%%%%

\begin{acknowledgments}
We thank the Natural Sciences and Engineering Research Council of
Canada for support of this research.
\end{acknowledgments}

%%%%%%%%%%%%%%%%%%%%%%%%%%%%%%%%%%%%%%%%%%%%%%%%%%%%%%%%%%%%%%%%%%%%%%%

\appendix

%%%%%%%%%%%%%%%%%%%%%%%%%%%%%%%%%%%%%%%%%%%%%%%%%%%%%%%%%%%%%%%%%%%%%%%

\section{\label{secapA} Numerical Implementation}

Here, we provide a route to compute the eigenvalues and overlap
integrals via Eq.\ (\ref{qket}).
We start by defining $N_\kappa$ and $N_\beta$ to be the basis-set
dimensions in the $Q$ and $P$ space, respectively, and
$N_T = N_\kappa + N_\beta$.
The probability of being in the $Q$ space, $P(t)$, is given by Eq.\
(\ref{totalpop2}).
In order to find $P(t)$, two sets of values are needed: the set
of eigenvalues $\{E_\gamma\}$, and the overlap integrals
$a_{\kappa,\gamma}$ between the zeroth-order states in $Q$ and the
exact eigenstates $\ket{\gamma}$.
The partitioning algorithm described below is ingenious in the
sense that it allows one to concentrate specifically on obtaining these
two sets of values. The method is well suited to small systems.

Beginning with Eq.\ (\ref{qket}), and using Eqs.\ (\ref{parameters}),
the algorithm is as follows:
\begin{enumerate}
 \item Choose a starting energy $E_{\gamma}^{i=0}$, with $i$
       corresponding to the $i$th iteration.
       In particular, one may choose an energy close to the
       zeroth-order energies.

 \item Take $E_{\gamma}^{i}$ from the last iteration, and compute
       ${\cal H}(E_{\gamma}^{i})$.

 \item Diagonalize ${\cal H}(E_{\gamma}^{i})$, and select one
       of its eigenvalues to be the next trial energy,
       $E_{\gamma}^{i+1}$.

 \item If $|E_{\gamma}^{i+1} - E_{\gamma}^{i}| \ncong 0$,
       go back to step 2.

 \item If $|E_{\gamma}^{i+1} - E_{\gamma}^{i}| \cong 0$,
       $E_{\gamma}^{i+1}$ becomes the eigenvalues $E_{\gamma}$, and its
       corresponding eigenvector, $\ket{D_\gamma}$, is proportional to
       $Q\ket{\gamma}$.

 \item Repeat steps 1-5 until all $N_T$ unique eigenvalues $E_{\gamma}$
       are obtained.
\end{enumerate}

In the process of diagonalizing the effective Hamiltonian, ${\cal H}$,
each eigenvector $\ket{D_\gamma}$ has been normalized to 1.
Therefore, the use of the algorithm leads to a loss of information
about $Q\ket{\gamma}$.
This makes necessary to also compute the constant of proportionality
between $Q\ket{\gamma}$ and $\ket{D_\gamma}$.
This is done by requiring that $\brak{\gamma}\ket{\gamma} = 1$ for
the full eigenvectors.
Thus, one can assert that
\beq
 Q\ket{\gamma} = C_\gamma\ket{D_\gamma} ,
 \label{qd}
\eeq
with $C_\gamma$ being the proportionality constant.
The problem then reduces to finding the $C_\gamma$ associated with
each $E_{\gamma}$.
This is accomplished by expressing $\brak{\gamma}\ket{\gamma}$ as
\begin{eqnarray}
 \brak{\gamma}\ket{\gamma} & = & \braket{\gamma}{Q}{\gamma} +
  \braket{\gamma}{P}{\gamma} \nonumber \\
 & = & \braket{\gamma}{Q^2}{\gamma} + \braket{\gamma}{P^2}{\gamma} ,
 \label{q2p2}
\end{eqnarray}
where
\beq
 \braket{\gamma}{Q^2}{\gamma} =
  |C_\gamma|^2 \brak{D_\gamma}\ket{D_\gamma} = |C_\gamma|^2 ,
 \label{qsquared}
\eeq
and, using Eq.\ (\ref{psolve}),
\begin{eqnarray}
 \braket{\gamma}{P^2}{\gamma} & = &
  \langle\gamma|QHP\left[ E_\gamma - PHP \right]^{-1}
  \nonumber \\
  & \times & \left[ E_\gamma - PHP \right]^{-1}PHQ|\gamma\rangle .
 \label{psquared}
\end{eqnarray}
The application of the spectral resolution of an operator, Eq.\
(\ref{specres}), to Eq.\ (\ref{psquared}) leads to
\begin{equation}
 \langle\gamma|P^2|\gamma\rangle = \sum_\beta \frac{
  \langle\gamma|Q H|\beta\rangle \langle\beta|H Q|\gamma\rangle}{
   \left( E_\gamma - \widehat{E}_\beta \right)^2} ,
\end{equation}
whereby, by making use of Eq.\ (\ref{qd}), one obtains
\begin{equation}
 \langle\gamma|P^2|\gamma\rangle = \left|C_\gamma\right|^2\sum_\beta
  \frac{ \langle D_\gamma|H|\beta\rangle
  \langle\beta|H|D_\gamma\rangle}{ \left( E_\gamma - \widehat{E}_\beta
   \right)^2} .
 \label{psquared2}
\end{equation}
Now $\langle\gamma|P^2|\gamma\rangle$ is easily computed by realizing
that
\begin{eqnarray}
 \langle D_\gamma|H|\beta\rangle & = & \sum_\kappa D_{\kappa\gamma}^*
  \langle\kappa|H|\beta\rangle \\
 & = & \sum_\kappa D_{\kappa\gamma}^* V(\kappa|\beta) .
\end{eqnarray}

The substitution of Eqs.\ (\ref{qsquared}) and (\ref{psquared2}) into
Eq.\ (\ref{q2p2}) yields
\begin{equation}
 \langle \gamma | \gamma \rangle = 1 = \left|C_\gamma\right|^2\left(
  1 + \sum_\beta \frac{\left|\sum_\kappa
  D_{\kappa\gamma}^*V(\kappa|\beta)\right|^2}
   {\left( E_\gamma - \widehat{E}_\beta \right)^2} \right) ,
 \label{eq_cgamma}
\end{equation}
from which one obtains the proportionality factor $|C_\gamma|$.

According to the procedure previously described, we can determine
$Q|\gamma\rangle$, given $|D_\gamma\rangle$, with the exception of a
constant phase factor.
Note that, in general, each proportionality factor, $C_\gamma$, can be
written as $|C_\gamma|e^{\theta_\gamma}$, where $\theta_\gamma$ is a
random phase.
However, this is not a problem since the results are independent of
any constant phase factor; as seen from Eq.\ (\ref{mkappa}), all
overlap integrals appear in pairs, $a_{\kappa',\gamma}
a_{\kappa,\gamma}^*$, which can be expressed as
\begin{eqnarray}
 \brak{\kappa}\ket{\gamma}\brak{\gamma}\ket{\kappa'} & = &
  \braket{\kappa}{e^{i\theta_\gamma}}{\gamma}
  \braket{\gamma}{e^{-i\theta_\gamma}}{\kappa'} \nonumber \\
  & = & \brak{\kappa}\ket{\gamma}\brak{\gamma}\ket{\kappa'} .
\end{eqnarray}

%%%%%%%%%%%%%%%%%%%%%%%%%%%%%%%%%%%%%%%%%%%%%%%%%%%%%%%%%%%%%%%%%%%%%%%
%%%%%%%%%%%%%%%%%%%%%%%%%%%%%%%%%%%%%%%%%%%%%%%%%%%%%%%%%%%%%%%%%%%%%%%


\begin{thebibliography}
\bibitem{PSOS}
 The surface of section has been computed in the standard way, i.e., by following
 each trajectory, and noting $R_1$ and $P_1$ each time that
 $R_2$ crosses the surface $R_2 = R_2^0$ with $P_2 >0$.
\bibitem{note2}
Perturbed trajectories were designed in the traditional manner,
modifying
 slightly $R_1$ ($R_1 \to R_1^\delta = R_1 + \delta$) and adjusting
$P_2$ to restore the original energy.
\end{thebibliography}
\end{document}